\documentclass[epjCONF]{svjour}
\usepackage{graphics}
\usepackage[varg]{txfonts} 
\usepackage[latin1]{inputenc}
\usepackage[]{natbib}  
\session-title{Conference Title, to be filled}
\begin{document}
%
\title{Towards age/rotation/magnetic activity relation with seismology
}
\author{Savita Mathur\thanks{\email{smathur@spacescience.org}} }
\institute{Space Science Institute, 4750 Walnut street Suite 205, Boulder, CO 80301, USA}
\abstract{
The knowledge of stellar ages directly impacts the characterization of a planetary system as it puts strong constraints on the moment when the system was born. Unfortunately, the determination of precise stellar ages is a very difficult task. Different methods can be used to do so (based on isochrones or chemical element abundances) but they usually provide large uncertainties. 
During its evolution a star goes through processes leading to loss of angular momentum but also changes in its magnetic activity. Building rotation, magnetic, age relations would be an asset to infer stellar ages model independently. Several attempts to build empirical relations between rotation and age (namely gyrochronology) were made with a focus on cluster stars where the age determination is easier and for young stars on the main sequence. For field stars, we can now take advantage of high-precision photometric observations where we can perform asteroseismic analyses to improve the accuracy of stellar ages. Furthermore, the variability in the light curves allow us to put strong constraints on the stellar rotation and magnetic activity. By combining these precise measurements, we are on the way of understanding and improving relations between magnetic activity, rotation, and age, in particular at different stages of stellar evolution. I will review the status on gyrochronology relationships based on observations of young cluster stars. Then I will focus on solar-like stars and describe the inferences on stellar ages, rotation, and magnetism that can be provided by high-quality photometric observations such as the ones of the {\it Kepler} mission, in particular through asteroseismic analyses. 
} 
\maketitle
\section{Introduction}
\label{intro}

The determination of stellar ages has been an important study in the past as they allow us to study the temporal evolution of different stellar parameters (such as the radius, the rotation period, the magnetic field) and to understand the physical processes in play to reach different stages of the life of a star with different values for these quantities. The physical processes include transport of angular momentum, mass loss, evolution of the magnetic field etc. 

The knowledge of stellar ages has even broader impacts in the field of astrophysics. Indeed they are important for planetary system studies as the host star and the planets form approximately at the same time.  Stellar ages are also important to understand star formation and galactic evolution.

Unfortunately, stellar ages are difficult to determine, specially for low-mass field stars. For coeval stars that belong to the same cluster, it is easier to determine the age as the stars also have similar metallicities. Another issue comes from the fact that some stellar phases evolve more or less rapidly depending on the mass of the star.

Different techniques can be used to measure stellar ages. Stars in open clusters represent an excellent opportunity to calibrate ages by fitting isochrones in the H-R Diagram. Indeed as they are coeval and with similar chemical composition precise ages can be determined. The same applies for stars in binary systems. Lithium abundance has also been used to determine ages for low-mass stars as it is easily destroyed in stars and is produced in special conditions. So a detection of Li in cool stars would indicate that the star is young. For massive stars, things seem to be more complicated as we would expect to observe high Li abundance but low abundances have been observed in massive stars. Finally, there also seems to exist a bi-modality depending on the effective temperature, which means that Lithium abundance can only help define a lower limit for seller ages. 

For field stars, isochrone fitting and asteroseismology are most commonly used. Finally, since rotation and magnetic activity change with the evolution of the star, gyrochronology and activity-age relations have been the center of many studies so that ages can be determined model independently. 
For a more detailed description of the stellar ads determination, I refer the reader to the review by \cite{2010ARA&A..48..581S}.

\begin{figure}[h]
\begin{center}
\resizebox{0.5\columnwidth}{!}{
\includegraphics{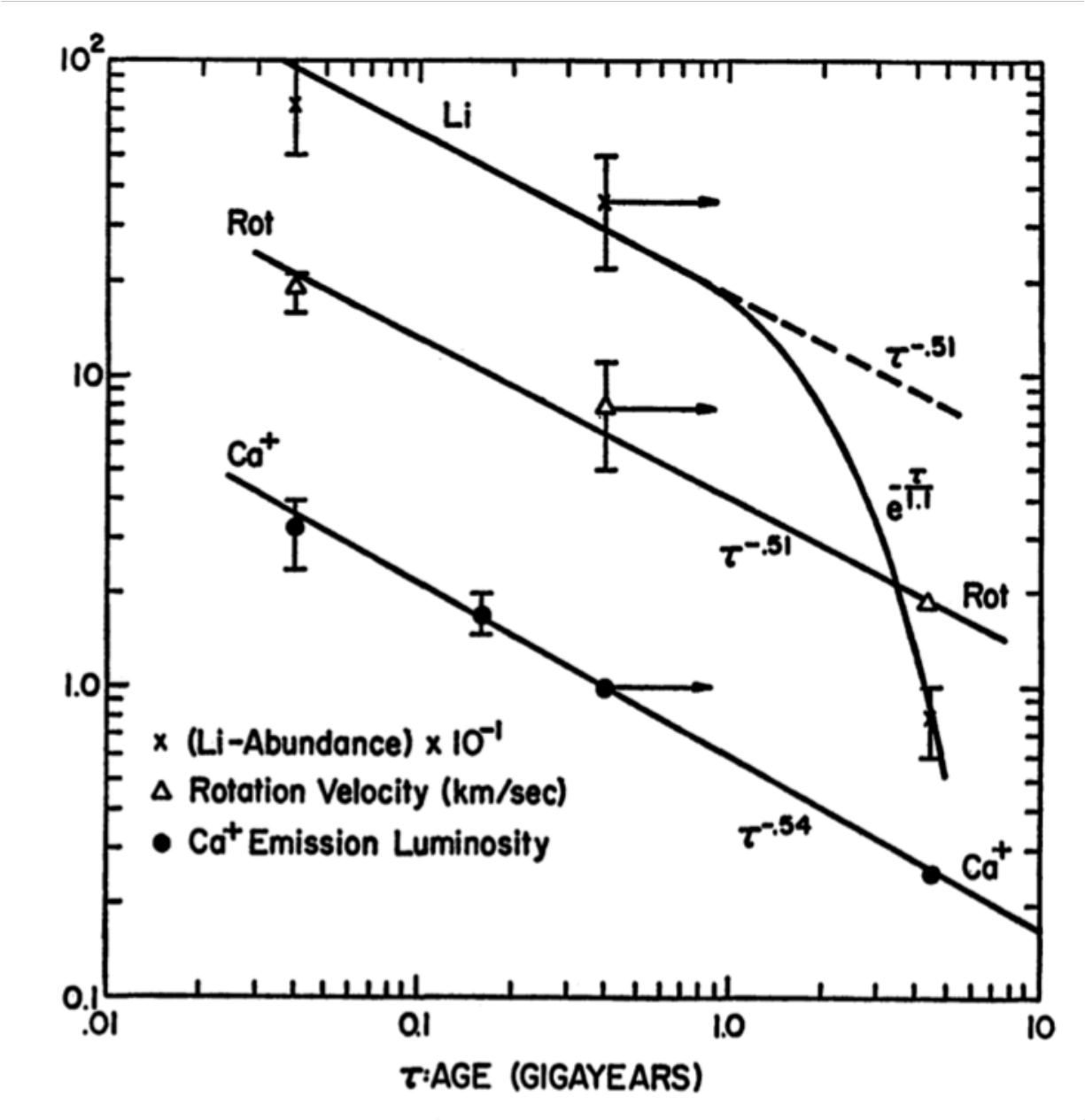}}
\caption{Lithium abundance, rotation velocity, and Ca emission for the Sun, Hyades stars and Ursa Major stars illustrating the $P_{\rm rot} \propto \sqrt{\tau}$ relation \citep[extracted from][]{1972ApJ...171..565S}.}
\label{fig:1}       
\end{center}
\end{figure}

Here I am going to briefly review relationships that classical observations of rotation and magnetic activity allowed us to derive. In the early 70s, \cite{1972ApJ...171..565S} studied the rotation, Calcium emission, and Lithium abundance for stars in three young clusters. Fig.\ref{fig:1} illustrates these observations as a function of the cluster ages. This allowed the author to derive the following relation: rotation period is inversely proportional to the square root of the stellar age. Later different theoretical studies were done to understand the transport of angular momentum \citep[e.g.][]{1988ApJ...333..236K,1991ApJ...376..204M,2013ApJ...776...67V,2014ApJ...788...93C}. \cite{1967ApJ...150..551K} had shown that there should also be a dependence with the stellar mass. Indeed stars more massive than 1.25\,M$_\odot$ seem to rotate fast. With their thin convective zones, these stars do not undergo magnetic braking and are less spun down than less massive stars. This is called the Kraft break. \cite{2014ApJ...780..159E} studied the rotation-mass-age relation.

Later, \cite{2003ApJ...586..464B,2007ApJ...669.1167B} added a dependency with the color, B-V, to the age-rotation relation. They measured the rotation periods of stars belonging to none young clusters with ages from 30 to 600\,Myr, as well as young and old field stars observed by the Mount Wilson observatory (with ages up to 10\,Gyr). They point out two branches in the relation between P$_{\rm rot}$ and B-V that evolve with time. 

\noindent The ``C branch'' corresponds to fast rotators that are fully convective. They do not have large-scale dynamo and are governed by small-scale convective turbulent field. These stars are mostly observed in young open clusters.

\noindent The ``I branch'' is constituted of slow rotators that have an interface dynamo. These stars have a large-scale interface magnetic field. As they have a deeper convective zone there is a coupling between the convective zone, the radiative zone and the exterior of the star. This branch is more and more populated as the cluster is older.

Regarding the age-activity relationship, \cite{2008ApJ...687.1264M} studied the chromospheric index based on CaHK observations ($R'_{HK}$) in young clusters with known ages and field stars with isochronal ages. They observed a drastic decrease of the magnetic index with time and derived a relation between $R'_{HK}$ and age. They also added a dependence with B-V that was also studied by \cite{2002AN....323..357S}.

More recently \cite{2013A&A...551L...8P} extended this study to a much larger sample (more than 500 field stars and several hundreds cluster stars) of stars with measurements of chromospheric activity indexes and ages from the Geneva-Copenhagen Survey. From this work, it seems that after 1.5\,Gyr, the chromospheric index does not evolve much with time.

\section{Photometric observations}

Photometric data from missions such as CoRoT \citep{2006cosp...36.3749B} or {\it Kepler} \citep{2010Sci...327..977B} contain a wealth of information like pulsations, rotation, and magnetic activity. The last ten years asteroseismic studies have been performed on a large number of targets of the CoRoT and {\it Kepler} missions  \citep[e.g.][]{2011A&A...534A...6C,2012A&A...543A..54A,2013A&A...549A..12M}. They have proved to considerably improve the determination of the mass, radius and age of the stars as well as the knowledge of their interiors \citep[e.g.][]{2012A&A...537A.111C,2012ApJ...749..152M,2013ApJ...763...49D,2014ApJS..210....1C,2014ApJS..214...27M}.

\subsection{Measuring rotation and magnetic activity}
\label{sec:2}

The presence of spots on the stellar surfaces creates a modulation in the light curve that is related to the surface rotation of the star. So for a star that is active (i.e. has spots) we can measure the surface rotation by computing a periodogram for instance \citep{2013A&A...557L..10N,2013A&A...560A...4R}. However, we must be very careful with this method as many times, instead of measuring the fundamental period we might detect a harmonic. Other techniques allow to be less biased by the harmonics such as the computation of the auto-correlation function \citep[ACF][]{2014ApJS..211...24M} or by performing a time-frequency analysis  \citep{2014A&A...562A.124M,2014MNRAS.441.2744V}. 

Since the measurement of rotation is tightly linked to presence of spots on the stars, we can also define photometric indexes for magnetic activity. Different magnetic indexes based on photometric data can be found in the literature \citep{2010Sci...329.1032G,2011ApJ...732L...5C,2013ApJ...769...37B}. They basically consist in measuring the standard deviation of the time series. Recently \cite{2014A&A...572A..34G} and \cite{2014A&A...562A.124M}  showed that we can define magnetic indexes based on the knowledge of the surface rotation of the star. This index $S_{\rm ph, k}$ is computed on subseries of length k times $P_{\rm rot}$. The final mean index is the mean value from all the sub series. This ensures that the index is related to the rotation timescales and hence to the evolution of the magnetic activity on the star. 

\subsection{Asteroseismic inputs}

Rotation lifts the degeneracy of the modes by leading to the observation of several components for modes of degree larger than 1. The separation between the components, called rotational splittings, is proportional to the rotation rate of the region probed by the mode. In addition, since the inclination angle between the line of sight and the rotation axis of the star is not 90 degrees like the Sun the splittings measured also depend on the this inclination angle like the v$\sin i$ obtained with spectroscopy. In the extreme case of a pole on observation the modes observed are not sensitive to rotation. Rotational splittings have been measured for thousands of modes in the Sun allowing to have a very good knowledge of the internal rotational profile down to 0.2\,R$_{\odot}$. To go deeper, we need the detection of splittings in gravity modes that probe the solar core. So far only asymptotic signature of the solar gravity modes has been detected \citep{2007Sci...316.1591G}. However the study of subgiants and red giants has been very successful with the measurement of rotational splittings of mixed  modes that are sensitive to both the stellar envelope and core.  These studies allowed in particular to discover that the core of these stars rotate faster than the surface \citep{2012Natur.481...55B,2012ApJ...756...19D,2014A&A...564A..27D,2012A&A...548A..10M}.     


Acoustic modes are affected by the magnetic field of the star. Indeed for the Sun it has been observed \citep{2009A&A...504L...1S,2013JPhCS.440a2020G} that when magnetic activity increases the modes have lower amplitudes and higher frequencies. The CoRoT mission allowed us to observe the same behavior in the solar-like star HD49933 \citep{2010Sci...329.1032G,2011A&A...530A.127S} by applying the similar technique \citep{2010A&A...511A..46M}. Study of other solar-like stars observed by CoRoT were not conclusive \citep{2013A&A...550A..32M}.  \cite{2011ApJ...732L...5C} studied the detection of acoustic modes in 2000 solar-like stars and showed a correlation between the non detection of modes and the magnetic activity level of the stars.

\section{Results from the {\it Kepler} mission}

Several studies on rotation, magnetic activity, and rotation-activity-age relationship have been done with the {\it Kepler} data and I will present here a few highlights. 

\subsection{Clusters}
\begin{figure}[h]
\begin{center}
\resizebox{0.5\columnwidth}{!}{
\includegraphics{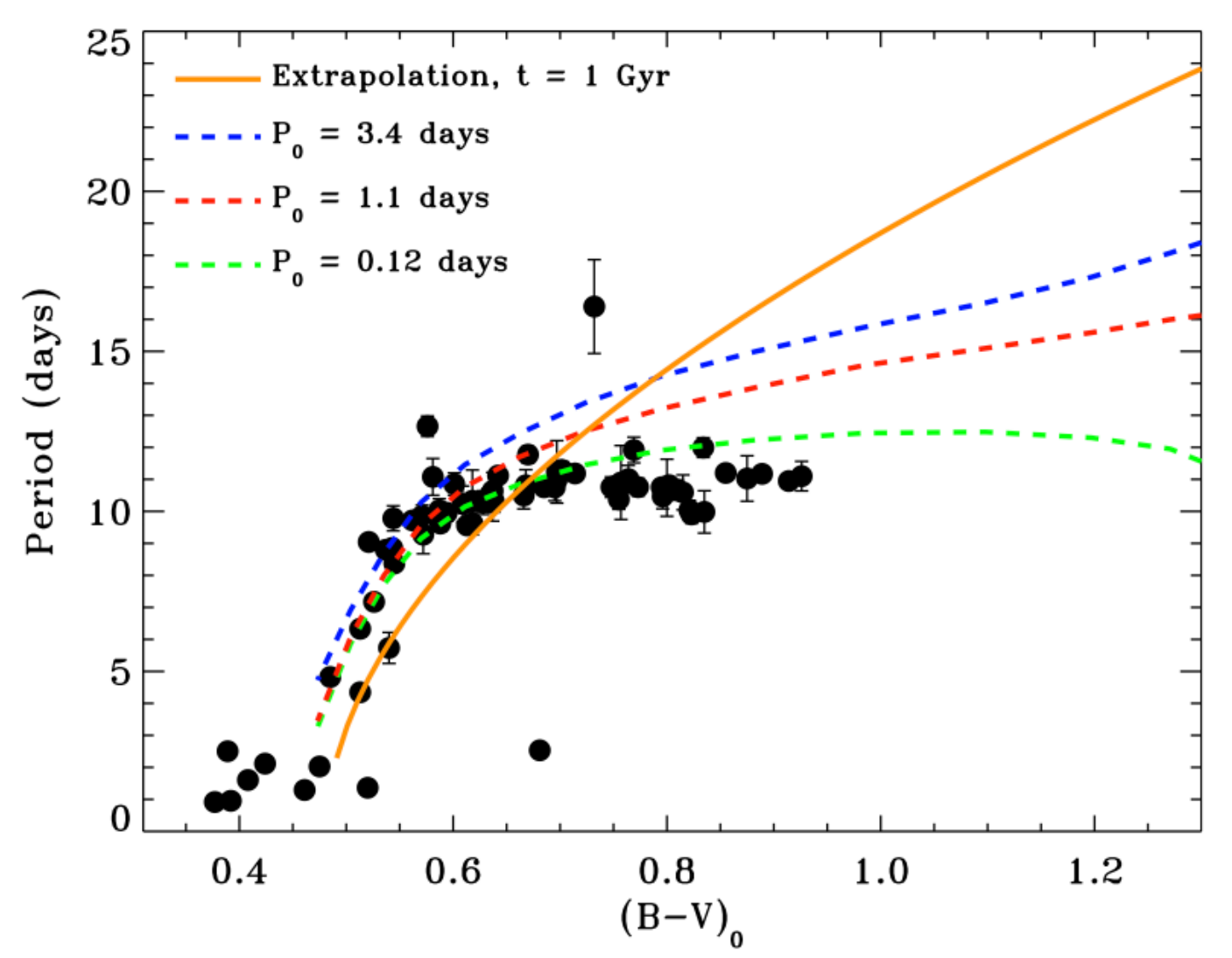}}
\caption{Rotation period vs B-V diagram for 71 FGK candidate single members of NGC 6811 with the extrapolation to 1\,Gyr of the Skumanich law (orange line) and rotational isochrones for 1\,Gyr using the relation from \cite{2010ApJ...722..222B} (dashed line) for different initial periods at the ZAMS. Extracted from \cite{2011ApJ...733..115M}.}
\label{fig:3}       
\end{center}
\end{figure}

Four clusters have been observed by the primary {\it Kepler} mission (NGC6866. 6811, 6819, 6791) with respective ages of 0.5, 1, 2.5, and 9\,Gyr. \cite{2011ApJ...733..115M} analyzed four quarters of 71 {\it Kepler} stars belonging to NGC6811  and computed the periodograms to measure the surface rotation periods. Fig.\ref{fig:3} shows the rotation periods as a function of the color B-V of the stars and compared the relationship with existing ones. In particular, they extrapolated the Skumanich spin down law that rotation period $P_{\rm rot} \propto \sqrt{t}$ \citep{1972ApJ...171..565S} computed for young clusters and found that this cannot reproduce the observation of the 1\,Gyr old cluster NGC6811 while the relation from \cite{2010ApJ...722..222B} seems to be more compatible.

\subsection{Field stars}

\cite{2014ApJS..211...24M} applied the ACF method to 34,030 stars observed for three years by the {\it Kepler} mission. By comparing the Period-effective temperature relationships from \cite{2007ApJ...669.1167B} and \cite{2008ApJ...687.1264M}, they conclude that there seems to be few stars older than 4.5\,Gyr and many stars younger than 1\,Gyr. However, all these stars have different masses and different evolutionary stages and having the Kraft break in mind all these stars should not be considered together.

This kind of study can be deepened for stars where asteroseismic analyses are available as asteroseismology can provide more accurate ages, masses, and radii as well as evolutionary stages. \cite{2014A&A...572A..34G} studied 540 solar-like stars where acoustic modes have been detected and found reliable surface rotation periods for 300 stars by combining ACF and time-frequency analyses. Since asteroseismic parameters are available for these stars, they divided their sample into three categories: hot stars, cool stars, and subgiants. They found that stars more massive than 1.2\,M$_{\odot}$ rotate in average faster than cooler or more evolved stars, which agrees with the Kraft break. They also fitted the rotation-age relationship for a subsample of cool stars with very precise ages and found values very similar to \cite{2007ApJ...669.1167B} and \cite{2008ApJ...687.1264M}, suggesting that relationships for young cluster stars are still valid for old field stars.

\begin{figure}[h]
\begin{center}
\resizebox{0.5\columnwidth}{!}{
\includegraphics{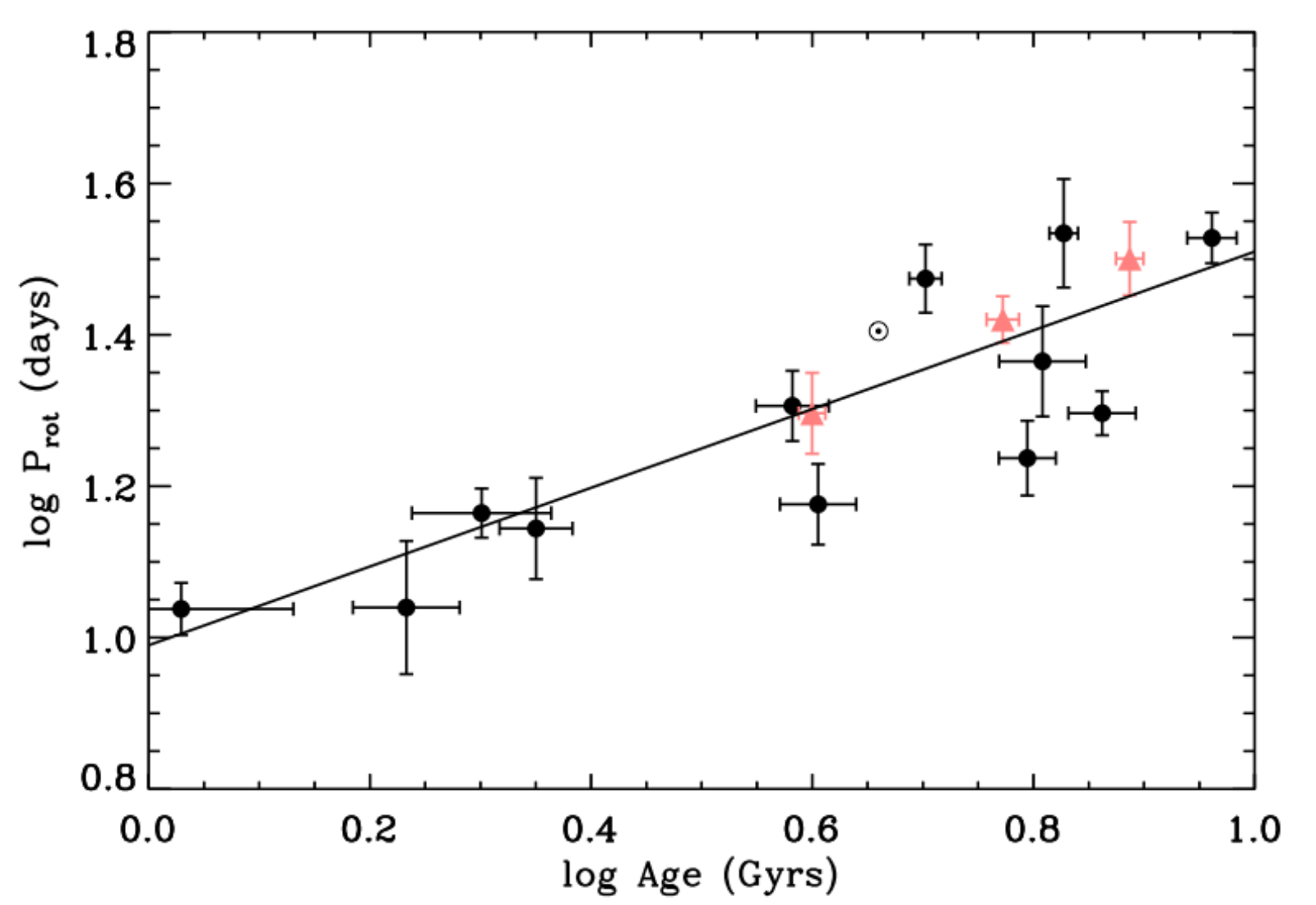}}
\caption{Surface rotation period as a function of age for a sample of 14 solar-like star with ages were determined with asteroseismic analyses \citep{2012ApJ...749..152M,2014ApJS..214...27M}. Extracted from \citet{2014A&A...572A..34G}.}
\label{fig:4}       
\end{center}
\end{figure}

\subsection{Solar analogs}

Among the sample of solar-like stars with seismology, we can also look for rotation of subsample of stars solar analogs or solar twins candidates, where solar analogs are stars with  0.9 $< M/M_{\odot} \le$ 1.1 and ``solar twin candidates'' refers to stars with 0.95 $< M/M_{\odot} \le$ 1.05 and rotation periods larger than 14 days.
\cite{2014ApJ...790L..23D} studied the rotation periods for eight seismic solar analogs and twins, and for five solar analogs based on the updated {\it Kepler} Input Catalog \citep{2014ApJS..211....2H}. They found that within the error bars, $P_{\rm rot}$ vs Age of seismic sample agree with previous relationships.

\subsection{Age-activity relations}

Relationships between magnetic activity and age are also interesting. Indeed the more evolved a star is the less active it becomes. \citep{2013MNRAS.433.3227K} observed 22 solar-like targets of the {\it Kepler} mission with the Nordic Optical Telescope. The CaHK observations seem to be anti-correlated with ages determined with asteroseismology. Later on, \cite{2014ApJS..214...27M} updated the ages with longer times series from {\it Kepler} and confirmed the anti-correlation.

%

\bibliographystyle{aa}  
\bibliography{/Users/Savita/Documents/BIBLIO_sav}

\end{document}